\documentclass{elsarticle}
\usepackage{amsmath}
\usepackage{graphicx, graphics}
\usepackage{amsfonts}
\usepackage{amssymb}
\usepackage{bm}

\newcommand{\op}{\varphi}
\newcommand{\ve}{\mathbf v}
\newcommand{\h}{\mathbf h}
\newcommand{\p}{\mathbf p}
\newcommand{\g}{\mathbf g}
\newcommand{\qu}{\mathbf q}
\newcommand{\bi}{\mathbf b}
\newcommand{\es}{\mathbf T_E}
\newcommand{\D}{\mathbf D}
\newcommand{\power}{\mathcal P} 
\newcommand{\ie}{e} 

\newcommand{\ent}{\eta}  
\newcommand{\fe}{\psi}  

\begin{document}
\begin{frontmatter}
\title{A phase field model for liquid-vapour phase transitions}
\author{V.Berti\corref{Cor1}}
\ead{berti@dm.unibo.it} 
\author{M.Fabrizio}
\ead{fabrizio@dm.unibo.it}
\author{D.Grandi}\ead{ grandi@dm.unibo.it}
\cortext[Cor1]{Corresponding author: phone number ++39-51-2094415, fax ++39-51-2094490}
\address{University of Bologna, Department of Mathemathics\\
Piazza di Porta S. Donato 5, I-40126, Bologna, Italy}

\begin{abstract}
We propose a model describing the liquid-vapour phase transition according to a phase-field approach. The model takes up the setting proposed in \cite{Fabrizio1}, where a phase field $\op$ is introduced whose equilibrium values $\op=0$ and $\op=1$ are associated to the liquid and vapour phases. The phase field obeys Ginzburg-Landau equation and enters the constitutive relation of the density, accounting for the sudden density jump occurring at the phase transition. In this paper we concern ourselves with the extension of this model to take into account the existence of the critical point in the coexistence line.
\end{abstract}


\end{frontmatter}

\section{Introduction}

In this paper an idea put forward in \cite{Fabrizio1} concerning a phase field description of water-vapour phase transition is developed. The main macroscopic manifestation of the liquid vapour phase transition (and, to a smaller extent, of the ice-water transition) is a sudden jump at a given pressure in the density of the substance. The starting assumption made in the cited paper is the decomposition of the density $\rho$ of the fluid into two contributions $\rho_0$ and $\rho_1$, such that
\begin{equation}\label{eqn:decomposition}
 \frac{1}{\rho}=\frac{1}{\rho_0}+\frac{1}{\rho_1}.
\end{equation}
 The contribution $\rho_0$ is taken as a function $\rho_0=\hat\rho_0(p,\theta)$ of the pressure $p$  and temperature $\theta$, while $\rho_1$ also hinges on an independent {phase field} $\op$. 
Such a decomposition entails the attainment of the Andrews curves with their characteristic plateaus. Otherwise, these would be obtained from the van der Waals state equation  supplemented by the Maxwell construction. In the cited approach, instead, it is the variation of the order parameter which gives rise to the horizontal piece of the isothermal lines (which prevents the relation between pressure and specific volume to be one-to-one). 
The additional phase field requires a further evolution equation (besides continuity, Navier-Stokes and heat equations) which is of course the time-dependent Ginzburg-Landau equation. We remark that in the classical Ginzburg-Landau approach to the liquid-vapour critical point, the order parameter is identified with the difference $\rho-\rho_c$ between the density $\rho$ and the critical density  $\rho_c$ (see {\it e.g.} \cite {Kadanoff, Chaikin}). Here, the order parameter is independent of the  density of the fluid and stands out from the constitutive decomposition (\ref{eqn:decomposition}) of the overall fluid density. \\
In this paper we concern ourselves with the modeling of the liquid-vapour phase transition along the same lines of \cite{Fabrizio1}, but also including the peculiarities due to the presence of the critical point. This aspect was not fully considered in the previous model, which makes sense specifically for temperatures below the critical temperature. As a consequence of the existence of the critical point, there always exists a sequence of equilibrium states which  connect in a {continuous} way the liquid and the vapour phase (see Fig. \ref{fig:critical-point}). Taking into account this kind of transformations, it is not possible  to assign a unique fixed value of $\op$ to the liquid and vapour phases anymore. We remark that, due to the existence of the critical point, it is possible to speak of distinct phases only on the coexistence curve. So it is necessary to consider a phase field which admits a continuous set of equilibrium values under different temperature and pressure conditions. Obviously this exhibits the second order (or continuous) character of the transition at the critical point. In  usual phase field models (see {\it e.g.} \cite{Fabrizio1, ABerti}) the homogeneous phases of a substance are associated to fixed discrete values of the order parameter (for example $\op=0$ and $\op=1$). Intermediate values are associated to the thin interfacial layers between two phases; in particular, they are associated, in stationary conditions, with non vanishing values of $\nabla\op$. This will not be true anymore if we assume a phase field which have a continuous set of equilibrium values associated to homogeneous phases, similarly to a Ginzburg-Landau order parameter in a strict sense (for second order phase transition). In this case intermediate values of the phase field can be associated both to interfacial layers and to homogeneous phases, close to the critical point.

\section{Model equations and thermodynamical consistence} \label{sec:model}

We start by setting the general thermodynamic framework of the model we are going to  propose. Our description is based on the introduction of an order parameter $\op$ connected with the density jump of the transition.  The equation for $\op$ is formulated as an additional balance equation to which is associated a corresponding balance of powers \cite{Fabrizio1, Fabrizio2, Gurtin}. As a consequence, a variation of $\rho_1$ is associated to a supplementary power flux in addition to the classical mechanical power $p\dot\rho/\rho$. 

We construct an evolutive model based on four balance equations for the fundamental thermomechanical fields of the fluid, that is: the phase field $\op$, the density $\rho$, the macroscopic velocity $\ve$ and the temperature $\theta$.\\
The thermomechanical state of the system is defined by the set of variables
\begin{equation}
 \sigma=(p,\theta,\op,\h),
\end{equation}
where $\h=\nabla\op$. We remark that we are considering a compressible fluid, both in its liquid  and in its vapour phase. In this situation, the choices of the pressure or the density $\rho$ as state variables are equivalent. From our point of view, it is more natural to choose the pressure as state variable, because it  appears explicitly in the Ginzburg-Landau equation for the order parameter.\\
We state the four balance equations of the model in the order: {order balance equation}, mass balance, momentum balance and energy balance
\begin{eqnarray}
 \rho k=\nabla\cdot\p,\label{eqn:order-bal}\\
 \dot\rho+\rho\nabla\cdot\ve=0,\\
 \rho\dot\ve=\nabla p+\nabla\cdot\es+\rho\bi,\\
 \rho\dot \ie= \power^i+\mathfrak h,
\end{eqnarray}
where:
\begin{itemize}
\item[-] the superimposed dot represents the material derivative of a field: $\dot \phi=\partial \phi/\partial t+\ve\cdot\nabla \phi$;
\item[-] $k$ is the internal order production and $\p$ the order flux. For the meaning of this terminology we refer to \cite{Fabrizio2}. The relation of these fields with the order parameter $\op$ is given in Eq. (\ref{eqn:kp});
\item[-] $\es$ is the extra stress (see Eq. (\ref{eqn:extrastress})), and $\bi$ the external force density;
 \item[-] $\ie$ is the internal energy density, $\power^i$ is the internal power density which is the sum of the internal mechanical power $\power^i_m$ associated to motion equation and the power $\power^i_{\op}$ associated to order balance; 
\item[-] $\mathfrak h$ is the internal thermal power density, which obeys its own balance in terms of heat flux $\qu$ and heat supply $r$:
\begin{equation}
 \mathfrak h=-\nabla\cdot \qu+r.
\end{equation}
\end{itemize}
The balance laws have to be completed by appropriate constitutive relations for $k,\p,\es,\rho$. 

The phase field equation suitable for the description of a phase transition takes the form of a Ginzburg-Landau equation:
\begin{equation}\label{eqn:GL}
 \rho\tau\dot\op=\kappa\nabla\cdot(\rho\nabla\op)-\rho f_\op(p,\theta,\op)
\end{equation}
where $\tau$ and $\kappa$ are positive parameters, $f$ is the Ginzburg-Landau potential and $f_\op=\partial f/\partial\op$. Hence the fields $k$ and $\p$ satisfy the constitutive equations
\begin{equation}\label{eqn:kp}
 k=\tau\dot\op+f_\op,\quad \p=\kappa\rho\nabla\op.
\end{equation}

Equation (\ref{eqn:GL}) allows us to write the balance of powers associated to order parameter by multiplying both members for $\dot\op$ and extracting a divergence contribution:
\begin{equation}
 \rho\tau\dot\op^2+\rho f_\op\dot\op+\rho\nabla\op\cdot\nabla(\kappa\dot\op)=\nabla\cdot(\kappa\rho\dot\op\nabla\op).
\end{equation}
We identify the first member with the internal power of the order parameter; using the identity
\begin{equation}
 \nabla\dot\op=\dot\h+\h\cdot\nabla\ve.
\end{equation}
we find useful to write the internal power as
\begin{equation}\label{eqm:P-op}
 \power_\op^i=\rho(\tau\dot\op^2+f_\op\dot\op+\kappa\h\cdot\dot\h+\kappa\h\cdot\D\h),
\end{equation}
where $\D=\frac12(\nabla\ve+\nabla\ve^T)$ is the rate of deformation tensor. 

The internal mechanical power $\power^i_m$ has the classical expression
\begin{equation}\label{eqm:Pm}
 \power_m^i=p\frac{\dot\rho}{\rho}+\es:\D
\end{equation}
and the extra-stress is assumed in the form
\begin{equation}\label{eqn:extrastress}
 \es=\mu(\op,\theta)\D-\kappa\rho\h\otimes\h.
\end{equation}
It includes a viscous contribution and a coupling term with the order parameter gradient $\h$ which is necessary to assure the thermodynamical consistence (see below). The viscosity $\mu(\theta,\op)$ is a function of the temperature and the phase which vanishes in the vapour and  gas phases.

The constitutive equation for $\rho$ is conveniently expressed in terms of the specific volume $\nu:=1/\rho$. We will assume a generic constitutive relation $\nu=\hat\nu(p,\theta,\op)$ \footnote{This is  different from the constitutive relation for density, pressure and temperature which is assumed in \cite{Fabrizio1}, which is rate dependent.}. The choice of the function $\hat\nu$ is strictly related to the function $f(p,\theta,\op)$ by the thermodynamical constraints and will be specified after we will have examined the consequences of the Second Law. We find it convenient to express the mechanical power (\ref{eqm:Pm}) in terms of the specific volume $\nu$  
\begin{equation}\label{eqm:Pm2}
 \power_m^i=-\rho p (\hat \nu_p\dot p+\hat \nu_\theta\dot\theta+\hat \nu_\op\dot\op)+\es:\D.
\end{equation}

Now we examine the constraints on the constitutive equations due to the Second Law, in the form of the Clausius-Duhem inequality. In terms of the free energy $\fe=e-\theta\ent$, the inequality can be expressed as
\begin{equation}
 \rho(\dot\fe+\eta\dot\theta)-\power_m^i-\power_\op^i+\frac{1}{\theta}\qu\cdot\g\leq 0,
\end{equation}
where $\g=\nabla\theta$. Substituting the expressions (\ref{eqm:P-op}) and (\ref{eqm:Pm2}) we obtain
\begin{eqnarray}\label{eqn:CD-ineq}
 (\fe_\theta+\ent+p\hat \nu_\theta)\dot\theta-\tau\dot\op^2+(\fe_\op-f_\op+p\hat \nu_\op)\dot\op+(\fe_p+p\hat \nu_p)\dot p+\\\nonumber
(\fe_\h-\kappa\h)\dot\h-(\frac{1}{\rho}\es+\kappa \h\otimes\h):\D+\frac{1}{\rho\theta}\qu\cdot\g\leq 0.
\end{eqnarray}
Firstly we observe that constitutive equation (\ref{eqn:extrastress}) makes the term proportional to $\D$ a non-negative definite term, provided that $\mu\geq0$. From inequality (\ref{eqn:CD-ineq}), it is plain the reason to include the contribution proportional to $\h\otimes\h$ in the extra stress tensor. 

In order to satisfy the Clausius-Duhem inequality in the most simple (non trivial) way, we add the following constitutive relations 
\begin{eqnarray}
 \ent&=&-\fe_\theta-p\hat \nu_\theta,\\
 \fe_\op&=&f_\op-p\hat \nu_\op,\\
 p\hat \nu_p&=&-\fe_p,\\
 \fe_\h&=&\kappa\h,\\
\qu&=&-k_0\g,\qquad k_0>0.
\end{eqnarray}
Then, the following expressions are easily obtained \footnote{Actually, one would obtain, for a given $f$, that $\fe(p,\theta,\op,\h)=\kappa \h^2/2+f(p,\theta,\op)+g(p,\theta)-pf_p$, but, as $g$ does not depend upon $\op$, we could rename $f+g$ as $f$, without modifying Ginzburg-Landau equation.}
\begin{eqnarray}
 \fe(p,\theta,\op,\h)&=&\frac{\kappa}{2}\h^2+f(p,\theta,\op)-pf_p(p,\theta,\op),\\
\hat \nu(p,\theta,\op)&=&f_p(p,\theta,\op),\label{eqn:hat-v}\label{eqn:v}\\
\ent(p,\theta,\op)&=&-f_\theta(p,\theta,\op).\label{eqn:ent}
\end{eqnarray}
 Note that
\begin{equation}\label{eqn:gibbs-fe}
 \Phi:=\frac{\kappa}{2}\h^2+f
\end{equation}
  takes the meaning of a specific Gibbs' free energy and $\psi=\Phi-p\Phi_p$.

Substituting the expressions of $\hat \nu$ and $\es$ in the internal power expression we obtain
\begin{equation}
 \power^i=\rho(\tau\dot\op^2+\dot\fe+\eta\dot\theta)+\mu||\D||^2.
\end{equation}
Then, the first law of thermodynamics
\begin{equation}
 \rho\dot\ie=-\nabla\cdot\qu+\rho r+\power^i
\end{equation}
gives us the heat equation
\begin{equation}
 -\rho\theta(f_\theta)^\cdot=\nabla\cdot(k_0\nabla\theta)+\rho r+\rho\tau\dot\op^2+\mu||\D||^2.
\end{equation}\\
We resume the set of equations describing the model
\begin{eqnarray}\label{eqn:system}
  \left\{\begin{array}{ll}
     \rho\tau\dot\op=\kappa\cdot\nabla(\rho\nabla\op)-\rho f_\op(p,\theta,\op)\\
      \dot\rho+\rho\nabla\cdot\ve=0\\
      \rho\dot\ve=\nabla p+\nabla\cdot(\mu(\op,\theta)\D-\kappa\rho\h\otimes\h)+\rho\bi\\
   -\rho\theta(f_\theta)^\cdot=\nabla\cdot(k_0\nabla\theta)+\rho r+\rho\tau\dot\op^2+\mu||\D||^2\\
 1/\rho=f_p(p,\theta,\op).
    \end{array}\right.
\end{eqnarray} 
The last equation is the constitutive equation for the density. It could be regarded as constitutive equation for the pressure if we would consider the density as a state variable.\\
The Clausius-Duhem inequality turns out to be very useful in establishing the coupling between the different equations. It has nothing to say about the choice of the function $f$ instead.\\
\\
\emph{Equilibrium properties.}\\
From the Ginzburg-Landau equation (\ref{eqn:GL}), the equilibrium value $\bar\op=\bar\op(p,\theta)$ of the order parameter is a solution  of the equation 
\begin{equation}\label{eqn:equilibrium}
f_\op(p,\theta,\bar\op)=0,
\end{equation}
which also should be a local minimum of $f$, for stability reasons. As the system has at most two coexisting phases, $f(p,\theta,\op)$ has to admit, as a function of $\op$,  at most two local minima $\bar\op_1,\,\bar\op_2$. On the \emph{coexistence line} 
\begin{equation}
p=p_0(\theta),\quad \theta<\theta_c,
\end{equation}
the two local minima have to be absolute minima, that is:
\begin{equation}
f(\bar\op_1, p_0(\theta),\theta)=f(\bar\op_2, p_0(\theta),\theta).
\end{equation}
In general, there is a strip-like region bordering the coexistence line in which both minima are present: one of them is an absolute minimum and represents the true equilibrium state in that region, while the other one is a relative minimum and represents a metastable state, which can be a superheated fluid or an undercooled vapour. Elsewhere, the order parameter has a unique equilibrium value, in particular for $\theta>\theta_c$ (or $p>p_c$).\\
The equilibrium Gibbs free energy (\ref{eqn:gibbs-fe}) for the homogeneous substance (that is for $\bm h=0$) in the $i$-phase is
\begin{equation}
\bar \Phi_i(p,\theta):=f(p,\theta,\bar\op_i(p,\theta))
\end{equation}
and the coexistence line is characterized by $\bar \Phi_1=\bar \Phi_2$.\\
The usual thermodynamics of phase equilibrium is then recovered. Firstly, we remark that, thanks to (\ref{eqn:equilibrium}), we have
\begin{equation}
\frac{\partial \bar \Phi_i(p,\theta)}{\partial p}=f_p(p,\theta,\op)|_{\op=\bar\op_i},\quad \frac{\partial \bar \Phi_i(p,\theta)}{\partial \theta}=f_\theta(p,\theta,\op)|_{\op=\bar\op_i}.
\end{equation}
So, the equilibrium \emph{equation of state} of the $i$-phase which follows from (\ref{eqn:v}) is represented by the usual relation
\begin{equation}\label{eqn:equil-vol}
\bar \nu_i(p,\theta)=\frac{\partial \bar \Phi_i(p,\theta)}{\partial p}.
\end{equation}
Similarly, from (\ref{eqn:ent}), the equilibrium entropy is given by 
\begin{equation}\label{eqn:equil-ent}
\bar\ent_i(p,\theta)=-\frac{\partial\bar \Phi_i(p,\theta)}{\partial \theta}.
\end{equation}
The specific volume jump due to phase change
\begin{equation}
\Delta \nu(\theta)=\bar \nu_2(p_0(\theta),\theta)-\bar \nu_1(p_0(\theta),\theta)
\end{equation}
is one of the most important parameters of the transition,  together with the latent heat
\begin{equation}
L(\theta)=\theta\Delta \ent=\theta[\,\bar\ent_2(p_0(\theta),\theta)-\bar\ent_1(p_0(\theta),\theta)\,].
\end{equation}
These quantities are related by the well known Clausius-Clapeyron equation
\begin{equation}
L=\theta\,p_0'(\theta)\Delta \nu,
\end{equation}
which can be easily obtained deriving both members of equilibrium condition
\begin{equation}
\bar \Phi_1(p_0(\theta),\theta)=\bar \Phi_2(p_0(\theta),\theta)
\end{equation}
with respect to $\theta$ and using (\ref{eqn:equil-vol}), (\ref{eqn:equil-ent}).\\

\section{The Ginzburg-Landau potential}

\begin{figure}[ht]
\includegraphics[scale=0.6]{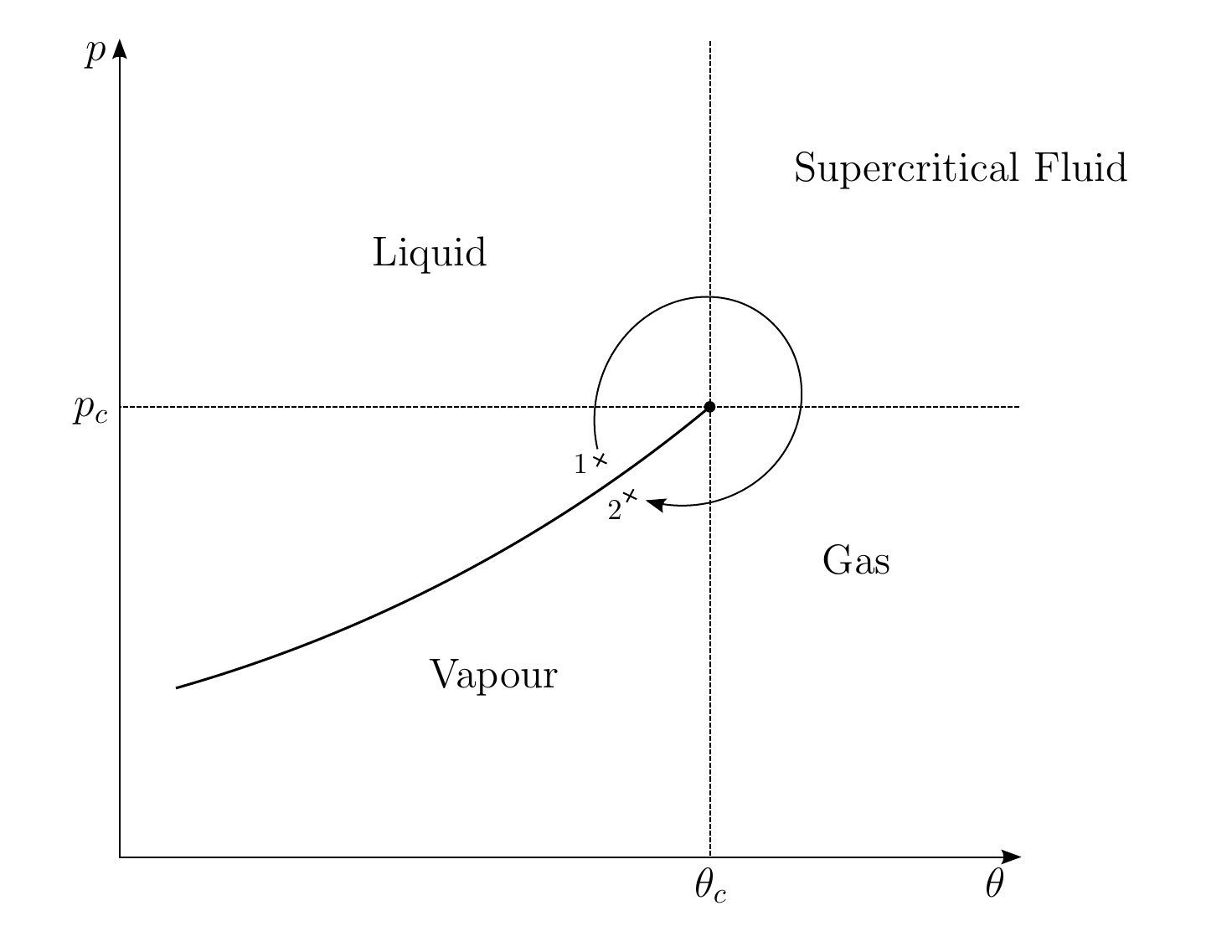}
\caption{The phase diagram near the liquid-vapour critical point. The transformation from state 1 to state 2 through the path indicated by the arrow does not involve discontinuities of any kind}
\label{fig:critical-point}
\end{figure}

We now turn to the choice of the Ginzburg-Landau potential $f$. The phase field approach to first order phase transitions involves an order parameter $\op$ which assumes a fixed value (say $\op=0$) in one of the two phases, and a different fixed value (say $\op=1$) in the other phase. The order parameter does not suffer jump discontinuities at phase boundaries, but varies smoothly across a interfacial layer from one equilibrium value to the other \cite{Gurtin, Brokate, Fremond}. This is obtained by subjecting the order parameter to a Ginzburg-Landau equation, with a potential $f(\op)$ having minima only at $\op=0,1$ for all values of the temperature and pressure (or any other relevant physical field). In the paper \cite{Fabrizio1} it was chosen a potential such that it had always two relative minima at $\op=0$ and $\op=1$. That minima were associated respectively to the liquid and vapour phases. This setting is not suited to model the phases in the neighborhood of the critical temperature. In fact, at the critical temperature $\theta_c$, the liquid and vapour phases merge in a unique phase which is sometimes named gas phase in a strict sense \footnote{We distinguish gas from vapour by saying that vapour can undergo a liquefaction transition at sufficiently high pressure, while gas does not.} (or \emph{supercritical fluid} in the region $\theta>\theta_c$, $p>p_c$). More precisely, at the critical point, a \emph{second order} phase transition occurs. We cannot associate a fixed value of the order parameter to the states lying on one side of the coexistence line, if this one have an end point. In fact, consider the transformation represented in Fig. \ref{fig:critical-point}. No phase transition occurs, so the order parameter has no jump discontinuity along that way. Therefore the order parameter should have a continuous set of equilibrium values. In particular, this rules out an approach based on three phases, liquid, vapour and gas, associated to fixed values of the order parameter (for example, 1, -1, 0): in that case, a physically inexistent first order phase transition between gas and liquid and between gas and vapour would take place somewhere in the phase diagram. 
The most direct generalization of the approach followed in \cite{Fabrizio1} is to assume  a Ginzburg-Landau potential with minimal points varying continuously with temperature at $\theta<\theta_c$. The two minima which are present in the coexistence region of liquid and vapour merge in a unique minimum $\op=0$ for $\theta\geq\theta_c$. The pressure will not move the position of the minima, but can make the order parameter jump from the liquid minimum to the vapour minimum and vice versa. So, in isothermal condition and inside the one phase region, the order parameter will be effectively constant and the Navier-Stokes equation is in fact decoupled from Ginzburg-Landau equation.\\
We start by considering the double well potential 
\begin{equation}
 F(x;u)=\frac{x^4}{4}-u^2\frac{x^2}{2},\quad u\geq 0.
\end{equation}
having two minima at $x=\pm u$ and a maximum at $x=0$. We next define the $C^1$ piecewise polynomial and odd function
\begin{equation}
 G(x;u)=\left\{
\begin{array}{lll}
\displaystyle \frac{x^3}{3}-u^2 x \quad &{\rm if}\ |x|\leq u\\
\\
      \displaystyle  -\frac{2}{3}sgn(x) u^3 \quad &{\rm if}\ |x|>u \\
      \end{array}
      \right.
\end{equation}

 Any linear combination $F(x;u)+h G(x;u)$, with a real coefficient $h$, is a function of $x$ having a minimum or a flexus in $x=\pm u$ and no other minimum. 

The Ginzburg-Landau free energy density is taken in the form
\begin{equation}
 f(p,\theta,\op)=f_0(p,\theta)+F(\op;\hat u(\theta))+\hat h(p,\theta)G(\op;\hat u(\theta)),
\end{equation}
where $\hat{h}$ is a function vanishing on the coexistence line, namely
$\hat{h}(p_0(\theta),\theta)=0$.
Then, the specific volume is given by
\begin{equation}
 \nu(p,\theta,\op)=f_p(p,\theta,\op)=(f_0)_p(p,\theta)+\hat h_p(p,\theta) G(\op;\hat u(\theta)).
\end{equation}
We identify the components of the decomposition (\ref{eqn:decomposition}) in the following way
\begin{equation}
 \frac{1}{\rho_0}=(f_0)_p(p,\theta),\quad \frac{1}{\rho_1}=\hat h_p(p,\theta) G(\op;\hat u(\theta)).
\end{equation}
Then, the difference in volume on the coexistence line, corresponding to the jumping in $\op$ from the minimum $\bar\op_1=u$ to the minimum $\bar\op_2=-u$, is
\begin{equation}
 \Delta \nu=\frac{4}{3 }\hat h_p(p_0(\theta),\theta)u^3.
\end{equation}
If we assume $\hat h(p,\theta)=p-p_0(\theta)$ for $\theta<\theta_c$, then the parameter $u$ is directly related to the measurable volume difference
\begin{equation}
 \hat u(\theta)=\left(\frac{3}{4}\Delta \nu(\theta)\right)^{1/3}.
\end{equation}
The liquid phase corresponds to the phase diagram region $\theta<\theta_c,\ p>p_0(\theta)$, that is, in the case we are considering, to $h>0$. For $h>0$ the absolute minimum is attained at $\op=u$, so, for this choice of $\hat h$, the liquid phase is associated to the right minimum, while the vapour phase is associated to the left minimum.\\
For $\theta>\theta_c$ we have $\Delta \nu=0$ and so $\hat u(\theta)=0$. We note that $G(x;0)=0$, so, for $\theta>\theta_c$ the Ginzburg-Landau potential is
\begin{equation}
f(p,\theta,\op)=f_0(p,\theta)+\frac{\op^4}{4}\quad (\theta>\theta_c),
\end{equation}
which has the only minimum $\op=0$. We associate that value to the gas phase.
\begin{figure}[h]
\includegraphics[width=11cm]{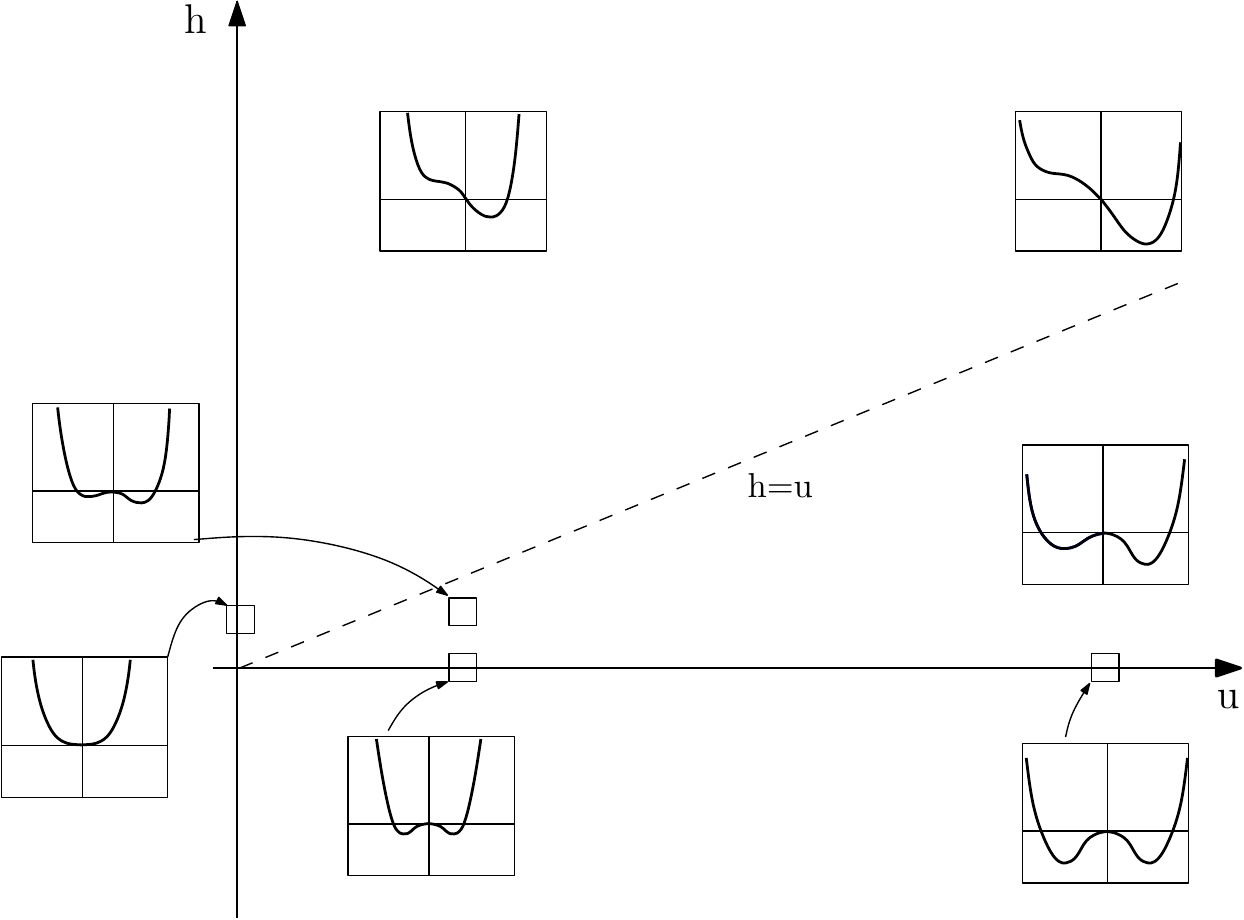}
\caption{Structure of the minima of the potential $f(p,\theta,\op)$ at varying $h>0,u>0$. In the region $h<0$ the structure is the same with the interchange of $\op$ with $-\op$.}
\label{fig:minimi-quart}
\end{figure}

On isothermal and homogeneous conditions (so that $\nabla\op=0$) and with externally prescribed homogeneous pressure field $p(t)$, the evolution of the volume is ruled by the equations
\begin{eqnarray}
 \tau\dot\op=-(F_\op(\op;u)+\hat h(p(t),\theta)G_\op(\op;u)),\\
\nu(t)=(f_0)_p(p(t),\theta)+\hat h_p(p(t),\theta)G(\op;u).
\end{eqnarray}
For $\tau\ll 1$, one can assume that at every instant the order parameter is equal to the equilibrium value $\bar\op(t)=\pm u$, so that the equilibrium equation of state is
\begin{equation}
 \nu(p,\theta)=(f_0)_p(p,\theta)+\hat h_p(p,\theta)G(\bar\op;u).
\end{equation}
The equilibrium value $\bar\op$ considered as a function of $p,\theta$, is not a single-valued function, since there is an hysteresis behaviour in the interval $|h|<u$. Considering only the stable states and discarding the metastable ones, amounts to take $\bar\op$ as 

\begin{equation}
 \bar\op(p,\theta)=\left\{
\begin{array}{lll}
u \quad &{\rm if}\ p>p_0(\theta)\\
\\
      \displaystyle  -u \quad &{\rm if}\  p<p_0(\theta)\\
      \end{array}
      \right.
\end{equation}

Assuming this function $\bar\op(p,\theta)$ the equilibrium isotherm at $\theta<\theta_c$ in the $\nu-p$ plane is
\begin{equation}
 \nu(p,\theta)=\left\{
\begin{array}{lll}
 (f_0)_p(p,\theta)-\frac{2u^3}{3}\quad &{\rm if}\ p>p_0(\theta)\\
\\
  (f_0)_p(p,\theta)+\frac{2u^3}{3} \quad &{\rm if}\  p<p_0(\theta)\\
      \end{array}
      \right.
\end{equation}
which has a flat piece at pressure $p_0(\theta)$ with length $\Delta \nu(\theta)=4u^3/3$. For pressures $p\neq p_0(\theta)$ the curve is determined essentially by $f_0(p,\theta)$. For $\theta\geq\theta_c$,  $\bar\op=0$ and $\nu=(f_0)_p$.

A difficulty in this model is given by a possible diverging contribution in the equilibrium entropy if $\Delta \nu(\theta)$ is singular at $\theta=\theta_c$, as, for example, if $\Delta \nu\sim|\theta-\theta_c|^\beta$, with $\beta<1$. In fact, consider the expression of the entropy
\begin{equation}
\ent(p,\theta,\op)=-f_\theta(p,\theta,\op)=\ent_0(p,\theta)+uu_\theta\op^2-h_\theta G(\op;u)+2huu_\theta\op,
\end{equation}
where $\ent_0(p,\theta):=\frac{\partial f_0}{\partial \theta}$. So, a diverging expression for $u_\theta$ would give rise to a diverging entropy at critical point.
 We can directly relate the equilibrium entropy $\bar\eta$ with $\Delta \nu(\theta)$. At the equilibrium values $\op=\pm u$, we have the entropies
\begin{equation}
\bar\ent_{\pm}=\ent_0+u_\theta u^3\pm \frac{2}{3}  h_\theta u^3 \pm 2hu^2u_\theta,
\end{equation}
or, remembering that $\Delta \nu(\theta)=4u^3/3$,
\begin{equation}
\bar\ent_{\pm}=\ent_0+\left(\frac{u}{4}\pm \frac{h}{2}\right)(\Delta \nu)_\theta\pm \frac{h_\theta}{2} \Delta \nu. 
\end{equation}
In particular, at the critical point $u=0$, $\Delta \nu=0$ and the entropy is given by
\begin{equation}
\bar\ent_{\pm}(p,\theta_c)=\ent_0(p,\theta_c)\pm \frac{1}{2}h(\theta_c,p)(\Delta \nu)_\theta. 
\end{equation}
Now, in general $(\Delta \nu)_\theta$ is a singular quantity at the critical point; for example, according to the Landau theory of the second order phase transitions, $\Delta \nu\propto |\theta-\theta_c|^{1/2}$. So, in this case we would obtain a divergent entropy at the critical temperature. In paper \cite{Fabrizio1} both equilibrium and non-equilibrium entropy function was a regular function of the temperature; nevertheless that model involved a temperature dependence of the  volume jump near critical point in the form     $\Delta\nu\sim|\theta-\theta_c|$. Furthermore, the latent heat does not approach to zero at critical point.\\
This shortcoming can be avoided in \emph{ad hoc} way assuming a cancelling divergence contribution in $\eta_0$ or following more closely the Landau theory of \emph{second order} phase transitions, in which the entropy remains a regular function at the critical point, while the equilibrium value of the order parameter has a square root singularity at critical point.\\
We suggest to consider the potential
 \begin{equation}
f(p,\theta,\op)=f_0(p,\theta)-a(u(\theta)+1)\ln(1-\op^2)-a\op^2-2h(p,\theta)\op,
\end{equation}
with $a>0$ and $u(\theta)$ an increasing function of $\theta$ such that
\begin{equation}
u(\theta)\in(-1,+\infty),\quad u(\theta_c)=0, \quad u(0^+)=-1,
\end{equation}
as, for example, in the following class of functions:
\begin{equation}
 u(\theta)=-\left[1-\left(\frac{\theta}{\theta_c}\right)^q\right]^{2\beta},\quad q>0,\ \beta>0.
\end{equation}
 
\begin{figure}[h]
 \includegraphics[width=10cm]{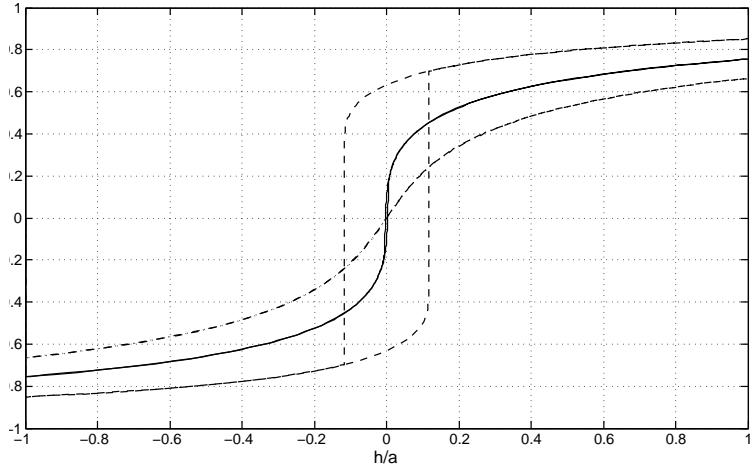}
\caption{Equilibrium solution $\bar\op$ as a function of $h/a$ for three values of $u$: $u=-0.4$ (dashed line), $u=0$ (continuous line), $u=0.4$ (dash-dotted line). For $u<0$ and $|h/a|<\bar h$ there are two minima, which have equal height at $h=0$; this gives rise to the characteristic hysteresis loop of the dashed line.}
\label{fig:op}
\end{figure}

As before, $h$ has to satisfy $h(p_0(\theta),\theta)=0$ and $h\lessgtr0$ for $p\lessgtr p_0(\theta)$.\\
The minima are solutions of the third order algebraic equation
\begin{equation}\label{eqn:eq-op}
 \op^3+u\op=\frac{h}{a}(1-\op^2).
\end{equation}
The same algebraic equation for the minima would be obtained (except for solutions $\bar\op\geq\pm1$) from the fourth order polynomial potential 
\begin{equation}
 F(\op)=\frac{\op^4}{4}+u\frac{\op^2}{2}-\frac{h}{a}\left(\op-\frac{\op^3}{3}\right),
\end{equation}
which is the classical form of the Ginzburg-Landau potential for second order phase transitions. The employment of the logarithmic potential (also used in different contexts, see for example \cite{FGM}) has the advantage that the only attainable minima are constrained in the interval $(-1,1)$. This would not be of great importance for the only study of critical region, where order parameter is small: $\op\ll1$. On the contrary, if one wants the phase field to be bounded under all the temperature and pressure conditions, the fourth order polynomial potential would not be suitable.
\begin{figure}[h]
\includegraphics[width=11cm]{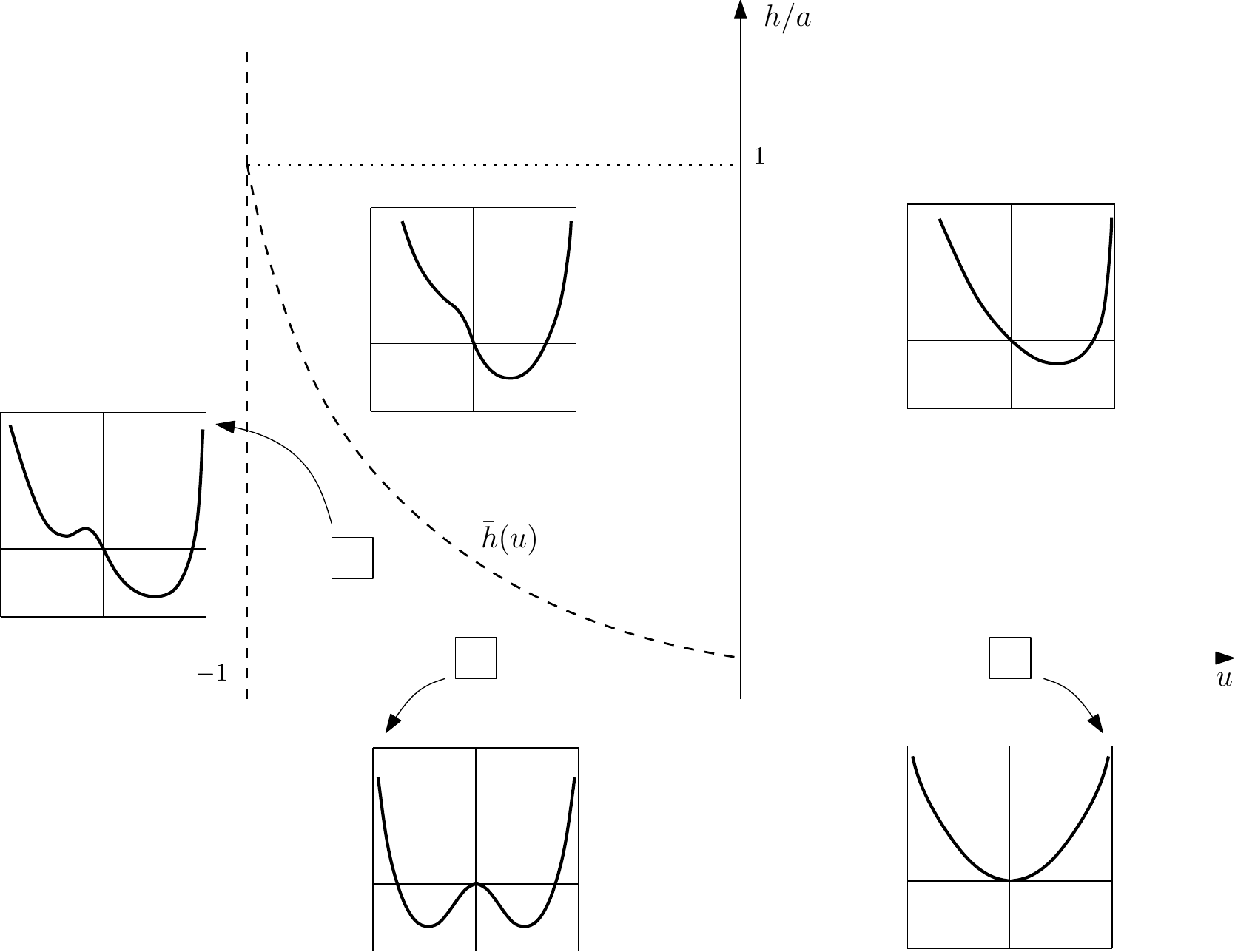}
\caption{Structure of the minima of the potential $f(p,\theta,\op)$ at varying $h>0,u>0$. In the region $h<0$ the structure is the same with the interchange of $\op$ with $-\op$.}
\label{fig:minimi-log}
\end{figure}
If $u<0$, there are two local minima $-1<\bar\op_1<0<\bar\op_2<1$ for $|h/a|< \bar h(u)$, while there is only one local minimum  $-1<\bar\op<1$ if $|h/a|\geq \bar h(u)$. (The function $\bar h(u)$ has not a simple expression, but $2|u/3|^{3/2}<\bar h<|u|$; see Fig. \ref{fig:minimi-log}). For  $u>0$, there is always only one minimum in $[-1,1]$. In particular, consider the case $h=0$; for $\theta<\theta_c$ (on the coexistence line), the minima are given by
\begin{equation}
 \bar\op_{\pm}=\pm |u|^{1/2}\in[-1,1],
\end{equation}
while for $\theta>\theta_c$ (on the critical isochore) the only minimum is $\bar\op=0$.\\
For $h\neq 0$, the minima have complicated algebraic expressions which are not very illuminating; we only remark that the absolute minimum satisfies the following estimate:
\begin{equation}
 |u|\frac{1-|u|}{|h|/a+|u|}\leq1-\bar\op^2\leq \frac{1-|u|}{|h|/a+1},\quad \textit{for } -1<u<0.
\end{equation}
In Fig. \ref{fig:op} it is plotted the value of $\bar\op$ in function of $h$ for three values of $u$.
The equation for volume is
\begin{equation}
 \nu=(f_0)_p-2h_p\op,
\end{equation}
which, owing to algebraic equation (\ref{eqn:eq-op}) for the equilibrium value $\bar\op$, implies the following state equation
\begin{equation}
 (f_{0p}-\nu)^3+\frac{2hh_p}{a}(f_{0p}-\nu)^2+4uh_p^2(f_{0p}-\nu)=\frac{8hh_p^3}{a}.
\end{equation}

The specific volume jump at the transition is
\begin{equation}
 \Delta \nu=4h_p|u|^{1/2}.
\end{equation}
If $u\sim(\theta/\theta_c-1)$, the quantity $\Delta \nu$ has the typical mean-field singularity $|\theta-\theta_c|^{1/2}$ near the critical point, needlessly to have  $u(\theta)$ also singular.\\ 
Finally, equilibrium entropy is given by
\begin{equation}
 \bar\eta=-(f_0)_\theta+ au'(\theta)\ln(1-\bar\op^2)+2h_\theta\bar\op.
\end{equation}
Now, $u(\theta)$ is a regular function of $\theta$, and so $\bar\eta$ is not diverging in the neighbourhood of the critical point.\\
With this choice of the Ginzburg-Landau potential, the equilibrium value (absolute minimum) of the order parameter $\bar\op$ is not constant in the inner of the regions of the phase diagram marked as 'liquid' and 'vapour'. In this sense, the Ginzburg-Landau equation is never completely decoupled from the Navier-Stokes equation. However, as
\begin{equation}
 1-\bar\op^2<\frac{a}{|h|}(1-|u|),
\end{equation}
the order parameter is approximatively $\pm1$ in the region $|h|\gg a$ (far from transition line) and/or $|u|\sim 1$ (well below the critical temperature).

\end{document}